\documentstyle{elsart}

\begin{document}

\begin{frontmatter}
\title{%
  Theory of the Luminescence Spectra of High-Density Electron-Hole
  Systems: 
  Crossover from Excitonic Bose-Einstein Condensation to Electron-Hole
  BCS state} 
\author[nara]{T.~J.~Inagaki\thanksref{e-mail}}
\author[nara]{M.~Aihara}
\author[nara]{A.~Takahashi}
\address[nara]{Graduate School of Materials Science,
  Nara Institute of Science and Technology,
  Ikoma, Nara 630-0101, Japan}
\thanks[e-mail]{E-mail:inagaki@ms.aist-nara.ac.jp}

\begin{abstract}
  We present a unified theory of luminescence spectra for highly
  excited semiconductors, which is applicable both to the
  electron-hole BCS state and to the exciton Bose-Einstein
  condensate. 
  The crossover behavior between electron-hole BCS state
  and exciton Bose-Einstein condensate clearly manifests itself in
  the calculated luminescence spectra.
  The analysis is based on the Bethe-Salpeter equation combined with
  the generalized random-phase-approximation, which enables us to
  consider the multiple Coulomb scattering and the quantum fluctuation 
  associated with the center-of-mass motion of electron-hole pairs. 
  In the crossover regime, the calculated spectra are essentially
  different from results obtained by the BCS-like mean-field theory
  and the interacting Boson model.
  In particular, it is found that the broad spectrum, arising from the
  recombination of electron-hole BCS state, splits into the P- and 
  P$\mbox{}_2$-luminescence bands with decreasing the particle
  density. 
  The dependence of these bands on the carrier density is in good
  agreement with experiments for highly excited semiconductors.
\end{abstract}
\begin{keyword}
  A. semiconductors, D. optical properties, D. phase transitions,
  D. electron-electron interactions
\end{keyword}

\end{frontmatter}

The issue of Bose-Einstein condensation (BEC) of excitons has received
considerable attention because of the Bosonic nature of excitons 
\cite{Moskalenko}.
Following the pioneering work by Keldysh {\it et al.} \cite{Keldysh},
the high-density electron-hole system with excitonic instability has
been extensively studied both theoretically \cite{Hanamura,Comte,Haug}
and experimentally \cite{Goto}. 
Recent developments in the experimental techniques make it possible to
observe the remarkable optical phenomena suggesting the spontaneous
generation of the macroscopic quantum coherence.
In particular, the anomalous exciton transport phenomena observed in 
$ {\rm Cu}_{\rm 2}{\rm O} $ \cite{Fortin} and 
$ {\rm BiI}_3 $ \cite{Karasawa} are manifestations of the
nondissipative rapid propagation of high-density excitons.
However, current understanding of these experiments still remains
controversial because of the complicated experimental situation due to
the finite life-time of excitons, phonon effects, spatial inhomogeneity
of the system, and so forth. 

We should remark that the macroscopic quantum state of high-density
excitons is qualitatively different from the BEC states of superfluid
$\mbox{}^4$He and atomic gases.
The system exhibits a crossover from the exciton BEC to the
electron-hole BCS state with increasing the excitation-light
intensity. 

Under the weak photoexcitation, the system can be described by the
interacting exciton model \cite{Hanamura} because the mean distance
between excitons is much larger than the exciton Bohr radius.
The system undergoes the exciton BEC at sufficiently low temperatures,
and the density of exciton with zero momentum is the order parameter
of the macroscopic quantum state.
When the photoexcitation is sufficiently strong, the bound
electron-hole pairs can no longer be regarded as pure Bosons because
the state-filling and the exchange effects, originated from the
Fermionic nature of electrons and holes, take essential part in the
problem.
In the high-density limit, the electron-hole pairs behave like
Cooper pairs, and the BCS-like energy gap at the Fermi level is
the order parameter of the macroscopic quantum state
\cite{Comte,Zimmermann,LK} which is essentially the same as the
excitonic phase \cite{Keldysh};
it is noted that no Bosonic nature exists above $ {\rm T}_{\rm c} $,
and the property of the system is totally different from the exciton
BEC state. 

Similar crossover phenomena have been discussed in a variety of
physical contexts, e.g., superconductivity \cite{NSR}, nuclear matter
\cite{Wada} and superfluid $\mbox{}^3$He \cite{Leggett}.
Recently, much attention has been focused on the BCS-BEC crossover
in connection with the unusual properties of the high-$T_{\rm c}$
cuprate superconductors.
In particular, it is observed that a pseudo-gap structure in the
normal-state density of states of underdoped cuprates persists almost
up to room temperature \cite{pseudo-gap}. 
A further unusual property of the high-$T_{\rm c}$ superconductor is
its extreme short coherence length (of the order of several lattice
constants) of Cooper pairs \cite{Uemura,Randeria}; this fact is in
contrast with the conventional superconductors where the Cooper pairs
are strongly overlapping in real space.
When we investigate this kind of crossover phenomena, the optically
generated electron-hole system has the marked advantage over the
superconducting materials, because the macroscopic quantum states can
directly be controlled by changing the intensity or the frequency of
excitation light.

In this article, we first present a unified theory of luminescence
spectra for highly excited electron-hole systems which is applicable
both to the electron-hole BCS state and to the exciton BEC state.
The earlier theories based on the interacting-Boson model or on the
BCS-like mean-field theory are not appropriate because the highly
photoexcited system is often in the intermediate state between the
electron-hole BCS state and the exciton BEC state.
In the present theory, the multiple Coulomb interaction between
electron-hole pairs is incorporated by numerically solving the
Bethe-Salpeter (BS) equation  \cite{Nakanishi} for electron-hole
pair correlation function.
This analysis is closely related to that in Ref.~\cite{Chu}, where
the absorption spectra for condensed exciton system are calculated by
properly taking into account the ladder diagram for vertex function.
In the present paper, the effect of the collective quantum
fluctuation associated with the center-of-mass motion of 
electron-hole pairs is incorporated by the generalized random-phase
approximation (RPA) \cite{Anderson}. 
The calculated spectra considerably deviate from results obtained
by the BCS-like mean-field analysis, and are essentially different
from results by the interacting Boson model.
The crossover from the exciton BEC to the electron-hole BCS state
is clearly found to manifest itself in the luminescence spectra.
In particular, the P$\mbox{}_2$-luminescence band,
which arises from the exciton recombination accompanied by the
excitation of another exciton from 1S to 2S state, continuously
changes to the Fermi-edge singularity with increasing the
electron-hole density. 
In the low carrier density, we find that the intensity of the
P$\mbox{}_2$-luminescence band predominates comparing to that of
the P-luminescence band, and this behavior is in excellent
agreement with experiments for various semiconductors
\cite{experiment1,experiment2}.
The present analysis also predicts the weak luminescence band due to
the recombination of electron-hole pairs which are generated by the
collective phase fluctuation.
These result are closely related to the recent noteworthy experiment
of ZnO thin films, in which the room temperature ultraviolet laser
emission is observed \cite{Kawasaki}.

We consider a electron-hole system in a direct-gap semiconductor,
which consists of the isotropic, nondegenerate parabolic conduction
and valence bands with identical electron and hole masses.
The repulsive interactions between electrons and between holes as well
as the electron-hole attractive interaction are taken into account.
The spin degrees of freedom is neglected to avoid the unnecessary
complication due to the biexciton state.

The spontaneous emission rate is expressed by the imaginary part of
the electron-hole pair correlation function as
$ I(\omega) = - 2 {\rm Im} G(\omega - \mu + i \gamma) $,
where $ \gamma $ and $ \mu $ are the exciton decay constant and the 
chemical potential of the electron-hole pairs, respectively.
$ G(\omega) $ is the Fourier transform of the electron-hole pair
correlation function
\begin{eqnarray}
  \label{eq:2-2}
  G(t) = - i \Theta(t)
  \sum_{\mbox{\boldmath \scriptsize $k$},
        \mbox{\boldmath \scriptsize $p$}}
  g_{\mbox{\boldmath \scriptsize $k$}}
  g_{\mbox{\boldmath \scriptsize $p$}}
  \langle
    c_{\mbox{\boldmath \scriptsize $k$}}^{\dagger}(0)
    d_{-\mbox{\boldmath \scriptsize $k$}}^{\dagger}(0)
    d_{-\mbox{\boldmath \scriptsize $p$}}(t)
    c_{\mbox{\boldmath \scriptsize $p$}}(t)
  \rangle ,
\end{eqnarray}
where $ g_{\mbox{\scriptsize \boldmath $k$}} $ is the radiation-matter
coupling constant; $ c_{\mbox{\boldmath \scriptsize $p$}} $ and
$ d_{\mbox{\boldmath \scriptsize $p$}} $ are the annihilation
operators for electrons and holes, respectively.
Here $ \langle \cdots \rangle $ indicates the expectation value with
respect to the electron-hole BCS state.
In order to consider the electron-hole pair correlation, let us
introduce the Bogoliubov transformation given by 
$
  c_{\mbox{\boldmath \scriptsize $k$}} = 
  u_{\mbox{\boldmath \scriptsize $k$}}
  \alpha_{\mbox{\boldmath \scriptsize $k$}} +
  v_{\mbox{\boldmath \scriptsize $k$}} 
  \beta_{-\mbox{\boldmath \scriptsize $k$}}^{\dagger}
$,  
$
  d_{-\mbox{\boldmath \scriptsize $k$}} = 
  u_{\mbox{\boldmath \scriptsize $k$}}
  \beta_{-\mbox{\boldmath \scriptsize $k$}} -
  v_{\mbox{\boldmath \scriptsize $k$}} 
  \alpha_{\mbox{\boldmath \scriptsize $k$}}^{\dagger},
$
where 
$ \alpha_{\mbox{\boldmath \scriptsize $k$}} $ and
$ \beta_{-\mbox{\boldmath \scriptsize $k$}} $ are the annihilation
operators for new quasiparticles (Bogolons).
The Bogoliubov parameters 
$ u_{\mbox{\boldmath \scriptsize $k$}} $ and 
$ v_{\mbox{\boldmath \scriptsize $k$}} $ are determined by solving the
BCS-like gap equation for electron-hole systems.
The BCS-like gap equation is reduced to the Wannier equation in the
low density limit, which implies that the equation gives the correct
result for the relative motion of electron-hole pairs both in the
high- and low-density limits \cite{NSR}.

In the following analysis, it is convenient to introduce a
two-component operator, 
$ \psi_{\mbox{\boldmath \scriptsize $k$}} =
  (\alpha_{\mbox{\boldmath \scriptsize $k$}},
   \beta_{-\mbox{\boldmath \scriptsize $k$}}^{\dagger}) $
and the Pauli spin matrices
$ \mbox{\boldmath $\tau$}_{\mu} $ ($ \mu = 1, 2, 3 $).
The spontaneous emission rate is expressed as
\begin{eqnarray}
  \label{eq:2-4}
  I(\omega) = \frac{2 \gamma I_{\rm coh}}{(\omega - \mu)^2 + \gamma^2}
  - \frac{1}{4}{\rm Im}
  \left[
    \sum_{\mbox{\boldmath \scriptsize $k$},
          \mbox{\boldmath \scriptsize $p$}}
    g_{\mbox{\boldmath \scriptsize $k$}}
    g_{\mbox{\boldmath \scriptsize $p$}}
    \mbox{\boldmath $K$}_{\mbox{\boldmath \scriptsize $k$}}
    \mbox{\boldmath ${\cal G}$}_{
      \mbox{\boldmath \scriptsize $k$},
      \mbox{\boldmath \scriptsize $p$}}
      (\omega - \mu + i \gamma)
    \mbox{\boldmath $K$}_{\mbox{\boldmath \scriptsize$p$}}
  \right] ,
\end{eqnarray}
where 
$ \mbox{\boldmath $K$}_{\mbox{\boldmath \scriptsize $k$}} = (
   u_{\mbox{\boldmath \scriptsize $k$}}^2 -
   v_{\mbox{\boldmath \scriptsize $k$}}^2, 
   i) $.
Here $ I_{\rm coh} $ is defined by
\begin{eqnarray}
  \label{eq:3-1a}
  I_{\rm coh} =
  \sum_{\mbox{\boldmath \scriptsize $k$},
        \mbox{\boldmath \scriptsize $p$}}
  g_{\mbox{\boldmath \scriptsize $k$}}
  g_{\mbox{\boldmath \scriptsize $p$}}
  u_{\mbox{\boldmath \scriptsize $k$}}
  v_{\mbox{\boldmath \scriptsize $k$}}
  u_{\mbox{\boldmath \scriptsize $p$}}
  v_{\mbox{\boldmath \scriptsize $p$}}
  \langle
  ( \psi_{\mbox{\boldmath \scriptsize $k$}}^{\dagger}
    \mbox{\boldmath $\tau$}_3
    \psi_{\mbox{\boldmath \scriptsize $k$}} )
  ( \psi_{\mbox{\boldmath \scriptsize $p$}}^{\dagger}
    \mbox{\boldmath $\tau$}_3
    \psi_{\mbox{\boldmath \scriptsize $p$}} )
  \rangle ,
\end{eqnarray}
and the $(\mu, \nu)$ component of 
$ \mbox{\boldmath ${\cal G}$}_{
    \mbox{\boldmath \scriptsize $k$},
    \mbox{\boldmath \scriptsize $p$}}
    (\omega) $
is defined by the Fourier transform of 
\begin{eqnarray}
  \label{eq:3-1b}
 \biggl(
  \mbox{\boldmath ${\cal G}$}_{
    \mbox{\boldmath \scriptsize $k$},
    \mbox{\boldmath \scriptsize $p$}} (t)
  \biggr)_{\mu, \nu} =
  - i \Theta(t) \langle
  ( \psi_{\mbox{\boldmath \scriptsize $k$}}^{\dagger}(0)
    \mbox{\boldmath $\tau$}_{\mu}
    \psi_{\mbox{\boldmath \scriptsize $k$}}(0) )
  ( \psi_{\mbox{\boldmath \scriptsize $p$}}^{\dagger}(t)
    \mbox{\boldmath $\tau$}_{\nu}
    \psi_{\mbox{\boldmath \scriptsize $p$}}(t) )
  \rangle ,
\end{eqnarray}
respectively.
The first term on the right-hand side of Eq.~(\ref{eq:2-4}) gives the
the sharp spectrum at the chemical potential of the electron-hole
pairs; this spectral component arises from the coherent spontaneous
emission from the macroscopic quantum state.
The electron-hole pair recombination does not generate Bogolons so
that the spectral width is merely determined by the lifetime of
electrons in the conduction band. 
The second term expresses the incoherent spontaneous emission rate
accompanied by the creation of Bogolons; this spectral component
reflects the various interesting features arising from the many-body
interaction in high-density electron-hole systems. 

Making use of the generalized RPA, we evaluate $ I_{\rm coh} $ in
Eq.~(\ref{eq:2-4}) as
\begin{eqnarray}
  \label{eq:2-5}
  I_{\rm coh} = 
  \left\{
    \sum_{\mbox{\boldmath \scriptsize $k$}}
    g_{\mbox{\boldmath \scriptsize $k$}}
    u_{\mbox{\boldmath \scriptsize $k$}}
    v_{\mbox{\boldmath \scriptsize $k$}}
  \right\}^2 
  + 2
    \sum_{\mbox{\boldmath \scriptsize $k$},
          \mbox{\boldmath \scriptsize $p$}}
    g_{\mbox{\boldmath \scriptsize $k$}}
    g_{\mbox{\boldmath \scriptsize $p$}}
    u_{\mbox{\boldmath \scriptsize $k$}}
    v_{\mbox{\boldmath \scriptsize $k$}}
    u_{\mbox{\boldmath \scriptsize $p$}}
    v_{\mbox{\boldmath \scriptsize $p$}}
    \int_{-\infty}^{+\infty}
      \frac{dz}{2 \pi {\rm i}}
      \frac{V_{
        \mbox{\boldmath \scriptsize $k$}
       -\mbox{\boldmath \scriptsize $p$}} (z)}
           {(z - E_{\mbox{\boldmath \scriptsize $k$}}
               - E_{\mbox{\boldmath \scriptsize $p$}})^2},
\end{eqnarray}
where 
$ V_{\mbox{\boldmath \scriptsize $q$}} (\omega) $ is the
screened Coulomb potential.
The single particle excitation energy (Bogolon energy) is given by
$ E_{\mbox{\boldmath \scriptsize $k$}} =
  \sqrt{\zeta_{\mbox{\boldmath \scriptsize $k$}}^2  +
        \Delta_{\mbox{\boldmath \scriptsize $k$}}^2 } $, 
where
$ \zeta_{\mbox{\boldmath \scriptsize $k$}} $ and 
$ \Delta_{\mbox{\boldmath \scriptsize $k$}} $ are the renormalized
energy of the electron-hole pair and the BCS gap, respectively.
The first term on the right-hand side of Eq.~(\ref{eq:2-5}) is
obtained by the BCS-like mean field approximation, and the second term
expresses the correction which comes from the collective quantum
fluctuations similar to the Anderson mode in superconductivity
\cite{Anderson}.

The function
$ \mbox{\boldmath ${\cal G}$}_{
    \mbox{\boldmath \scriptsize $k$}, 
    \mbox{\boldmath \scriptsize $p$}} (\omega) $,
which gives the intensity of the incoherent emission through
Eq.~(\ref{eq:2-4}), satisfies the following BS equation,
\begin{eqnarray}
  \label{eq:2-6}
  \mbox{\boldmath ${\cal G}$}_{
    \mbox{\boldmath \scriptsize $p$},
    \mbox{\boldmath \scriptsize $k$}} (\omega)
  (\omega \mbox{\boldmath $\tau$}_0
  + 2 E_{\mbox{\boldmath \scriptsize $k$}} \mbox{\boldmath $\tau$}_2) 
  - \sum_{\mbox{\boldmath \scriptsize $k$}'}
  V_{\mbox{\boldmath \scriptsize $k$}
    -\mbox{\boldmath \scriptsize $k$}'} (\omega)
  \mbox{\boldmath ${\cal G}$}_{
    \mbox{\boldmath \scriptsize $p$},
    \mbox{\boldmath \scriptsize $k$}'} (\omega)
  \left(
    C^{(0)}_{
      \mbox{\boldmath \scriptsize $k$},
      \mbox{\boldmath \scriptsize $k$}'}
    \mbox{\boldmath $\tau$}_2
  + i
    C^{(3)}_{
      \mbox{\boldmath \scriptsize $k$},
      \mbox{\boldmath \scriptsize $k$}'}
    \mbox{\boldmath $\tau$}_1
  \right)
\nonumber \\
  = \langle 
      (\psi_{\mbox{\boldmath \scriptsize $k$}}^{\dagger}(0)
       \mbox{\boldmath $\tau$}_{\mu}
       \psi_{\mbox{\boldmath \scriptsize $k$}}(0))
      (\psi_{\mbox{\boldmath \scriptsize $p$}}^{\dagger}(0)
       \mbox{\boldmath $\tau$}_{\nu}
       \psi_{\mbox{\boldmath \scriptsize $p$}}(0))
    \rangle ,
\end{eqnarray}
where
$ C^{(0)}_{
    \mbox{\boldmath \scriptsize $k$},
    \mbox{\boldmath \scriptsize $p$}} \equiv
 (u_{\mbox{\boldmath \scriptsize $k$}}
  u_{\mbox{\boldmath \scriptsize $p$}} +
  v_{\mbox{\boldmath \scriptsize $k$}}
  v_{\mbox{\boldmath \scriptsize $p$}})^2 $
and
$ C^{(3)}_{
    \mbox{\boldmath \scriptsize $k$},
    \mbox{\boldmath \scriptsize $p$}} \equiv
 (u_{\mbox{\boldmath \scriptsize $k$}}
  v_{\mbox{\boldmath \scriptsize $p$}} -
  v_{\mbox{\boldmath \scriptsize $k$}}
  u_{\mbox{\boldmath \scriptsize $p$}})^2 $
are the coherence factors and
$ \mbox{\boldmath $\tau$}_0 $ is the $ 2 \times 2 $ unit matrix.
This BS equation is obtained by linearizing the equation-of-motion for
the correlation function
$ \mbox{\boldmath ${\cal G}$}_{
    \mbox{\boldmath \scriptsize $k$},
    \mbox{\boldmath \scriptsize $p$}}(t) $.
The multiple Coulomb interaction between electrons and
holes is expressed by the second term on the right-hand side of 
Eq.~(\ref{eq:2-6}).
This term makes a significant contribution to the recombination of the
electron-hole pairs with nonzero total momentum, which is not taken
into account in the BCS-like mean-field theory.
The right-hand side of Eq.~(\ref{eq:2-6}) is rewritten as 
\begin{eqnarray}
  \label{eq:a1}
    \langle
      (\psi_{\mbox{\boldmath \scriptsize $k$}}^{\dagger}(0)
       \mbox{\boldmath $\tau$}_{\mu}
       \psi_{\mbox{\boldmath \scriptsize $k$}}(0))
      (\psi_{\mbox{\boldmath \scriptsize $p$}}^{\dagger}(0)
       \mbox{\boldmath $\tau$}_{\nu}
       \psi_{\mbox{\boldmath \scriptsize $p$}}(0))
    \rangle
\nonumber \\
  = (\mbox{\boldmath $\tau$}_0 + \mbox{\boldmath $\tau$}_2)
  \delta_{\mbox{\boldmath \scriptsize $k$},
          \mbox{\boldmath \scriptsize $p$}}
  - \frac{1}{2} \mbox{\boldmath $\tau$}_3
    \sum_{j=1}^{2} (-1)^{j+1}
    \langle
     (\psi_{\mbox{\boldmath \scriptsize $k$}}^{\dagger}(0)
      \mbox{\boldmath $\tau$}_{j}
      \psi_{\mbox{\boldmath \scriptsize $p$}}(0))
     (\psi_{\mbox{\boldmath \scriptsize $k$}}^{\dagger}(0)
      \mbox{\boldmath $\tau$}_{j}
      \psi_{\mbox{\boldmath \scriptsize $p$}}(0))
    \rangle .
\end{eqnarray}
The second term on the right-hand side represents the fluctuation
effect due to the collective excitations similar to the Anderson mode
in superconductivity.
A generalized RPA analysis gives the following expression,
\begin{eqnarray}
  \label{eq:a2}
    \sum_{j=1}^{2} (-1)^{j+1}
    \langle
     (\psi_{\mbox{\boldmath \scriptsize $k$}}^{\dagger}(0)
      \mbox{\boldmath $\tau$}_{j}
      \psi_{\mbox{\boldmath \scriptsize $p$}}(0))
     (\psi_{\mbox{\boldmath \scriptsize $k$}}^{\dagger}(0)
      \mbox{\boldmath $\tau$}_{j}
      \psi_{\mbox{\boldmath \scriptsize $p$}}(0))
    \rangle
    = 4
   \int_{-\infty}^{+\infty} \frac{dz}{2 \pi i}
   \frac{
    V_{
      \mbox{\boldmath \scriptsize $p$}
     -\mbox{\boldmath \scriptsize $k$}}(z)
    C^{(3)}_{
    \mbox{\boldmath \scriptsize $p$}, 
    \mbox{\boldmath \scriptsize $k$}}}
       {z^2 - (E_{\mbox{\boldmath \scriptsize $p$}} + 
               E_{\mbox{\boldmath \scriptsize $k$}})^2} .  
\nonumber  \\
\end{eqnarray}

As a first step in our numerical calculations, the gap equation is
iteratively solved to evaluate the renormalized energy of
electron-hole pair $ \zeta_{\mbox{\boldmath \scriptsize $k$}} $ and
the BCS gap function $ \Delta_{\mbox{\boldmath \scriptsize $k$}} $.
The incoherent luminescence spectrum is evaluated by numerically
calculating the eigenvalues and eigenvectors of the stability matrix
\cite{Ring} for Eq.~(\ref{eq:2-6}).
The quasi-static dielectric function given by the
single-plasmon-pole approximation is employed \cite{Haug,Zimmermann}.

The incoherent components of the luminescence spectra are shown in 
Figs.~\ref{fig:1}-\ref{fig:4}.
As a reference, we also present the luminescence spectra calculated by
the BCS-like mean-field theory and the generalized RPA analysis, in which
the fluctuation effect associated with the center-of-mass motion of
pairs is considered but the multiple Coulomb interaction is neglected.
We use the units where the exciton binding energy and the exciton Bohr
radius being unity. 
The dimensionless mean interparticle distance 
$ r_{\rm s} = (4 \pi n/3)^{-1/3} $ is introduced, where $ n $ is the
electron-hole density.
The exciton decay constant is chosen as $ \gamma = 0.03 $, unless
othewise stated.
The arrow in the figures represents the level of the chemical
potential of the electron-hole pair at which the coherent luminescence
line appears. 
With decreasing the particle density, the position of coherent
luminescence band approaches to $ \omega - E_{\rm g} = -1 $, where  
$ E_{\rm g} $ is the energy gap between the valence and conduction
bands. 
This behavior ensures that the present analysis reproduces the result
obtained by the interacting exciton model in the low density limit.

Fig.~\ref{fig:1} shows the luminescence spectrum in the high density
case ($ r_{\rm s} = 1.0 $).
The band renormalization effect due to the exchange interaction is
clearly observed. 
The peaked structure near the Fermi level 
($ \omega - E_{\rm g} \simeq 1 $) is the manifestation of the BCS-like
gap formation, which is strongly pronounced by the multiple Coulomb
interaction between electrons and holes; the electron-hole pairs with
nonzero total momentum take part in the recombination processes.
The luminescence band in $ \omega - E_{\rm g} > \mu $, which is missing in
the mean-field analysis, corresponds to the pair
annihilation process of the virtual Bogolons generated by the
collective phase fluctuation.

The luminescence spectra in the intermediate densities corresponding
to $ r_{\rm s} = 2.2 $ is shown in Fig.~\ref{fig:2}.
It is noted that the incoherent luminescence spectrum below the
chemical potential splits into two components:
the broad component below $ \omega - E_{\rm g} \simeq -1.7 $ and
the sharp one around $ \omega - E_{\rm g} \simeq -1.5 $.
The intensity of the broad component is approximately proportional to 
the square of the particle density;
its spectral position slightly shifts towards the shorter wavelength
with decreasing the particle density.
These behaviors suggest that the broad
component corresponds to the P-luminescence band
\cite{experiment1,experiment2,P-line}, which comes from the radiative
recombination of an exciton assisted by the dissociation of another
exciton.
On the other hand, the sharp component around 
$ \omega - E_{\rm g} \simeq -1.5 $ corresponds to the
P$\mbox{}_2$-luminescence band \cite{experiment1,experiment2}, which
arises from the radiative recombination of an exciton accompanied by
the excitation of another exciton from 1S to 2S state. 
In addition, a weak luminescence subband, called P$\mbox{}_3$, is
observed around $ \omega - E_{\rm g} \simeq -1.6 $;
this component comes from the exciton recombination assisted by
the excitation of another exciton from 1S to 3S state.
The peaked structure above the chemical potential originates from
the pair annihilation of Bogolons, which are generated by the
collective phase fluctuation from the electron-hole BCS state.
It is noted that we do not deal with a simple two-exciton problem, but
calculated P, P$\mbox{}_2$, and P$\mbox{}_3$ bands in the framework of
the many-body theory.
We should also recall that these spectral components cannot be
obtained by a BCS-like mean-field theory with the RPA correction.

The luminescence spectrum in a low electron-hole density state 
($ r_{\rm s} = 13 $) is shown in Fig.~\ref{fig:4}.
It is found that the P-luminescence band almost disappears and the
P$\mbox{}_2$ band predominates with decreasing the carrier density.
This behavior arises because the transition of another exciton to
the bound state (2S) is much stronger than that to the dissociated
states in low carrier density state.

We show in Fig,~\ref{fig:5} the comparison between the present theory
and the experiment for ZnO thin film at 77K \cite{experiment1}.
Here the exciton decay constant is chosen as $ \gamma = 0.1 $.
The calculated spectra are in good agreements with this experimant
in the following points: (1) the exciton luminescence line
predominates in the low density state.
(2) The intensity of the $ {\rm P}_2 $-luminescence line grows
superlinearly with increasing carrier density, and it prevails in
the intermediate density states.
(3) The P-luminescence line appears in higher densities and it
merges with the $ {\rm P}_2 $-luminescence line as the carrier density
increases.

Finally, it should be noted that the dependence of the line shape
and the spectral position on the electron-hole density cannot be
predicted by the well-known analysis for two-exciton problem
\cite{Kushida}.

In conclusion, we first present the many-body theory for luminescence
spectra which is applicable both to the exciton BEC state and to the
electron-hole BCS state. 
Our analysis, based on the BS equation combined with the generalized
RPA, properly describes the crossover behavior between the two
essentially different macroscopic quantum states. 

This work is partially supported by a Grant-in-Aid for Scientific
Research from the Ministry of Education, Science, Culture and Sports
of Japan. 



\begin{figure}[h]
\caption{%
  The luminescence spectrum in the high-density state
  ($ r_{\rm s} = 1.0 $) calculated by (a) the BCS-like mean-field
  theory, (b) the generalized RPA, and (c) the present theory.
}
\label{fig:1}
\end{figure}

\begin{figure}[h]
\caption{%
  The luminescence spectrum in the intermediate-density state
  ($ r_{\rm s} = 2.2 $) calculated by (a) the BCS-like mean-field
  theory, (b) the generalized RPA, and (c) the present theory.
}
\label{fig:2}
\end{figure}

\begin{figure}[h]
\caption{%
  The luminescence spectrum in the low-density state
  ($ r_{\rm s} = 13 $) calculated by (a) the BCS-like mean-field
  theory, (b) the generalized RPA, and (c) the present theory.
}
\label{fig:4}
\end{figure}

\begin{figure}[h]
\caption{%
  The luminescence spectrum for various density states:
  (a) the present theory, (b) experiment for ZnO thin film at 77K
  (from Ref.~\cite{experiment1}).
  Here Ex-line corresponds to the free exciton emission.
}
\label{fig:5}
\end{figure}

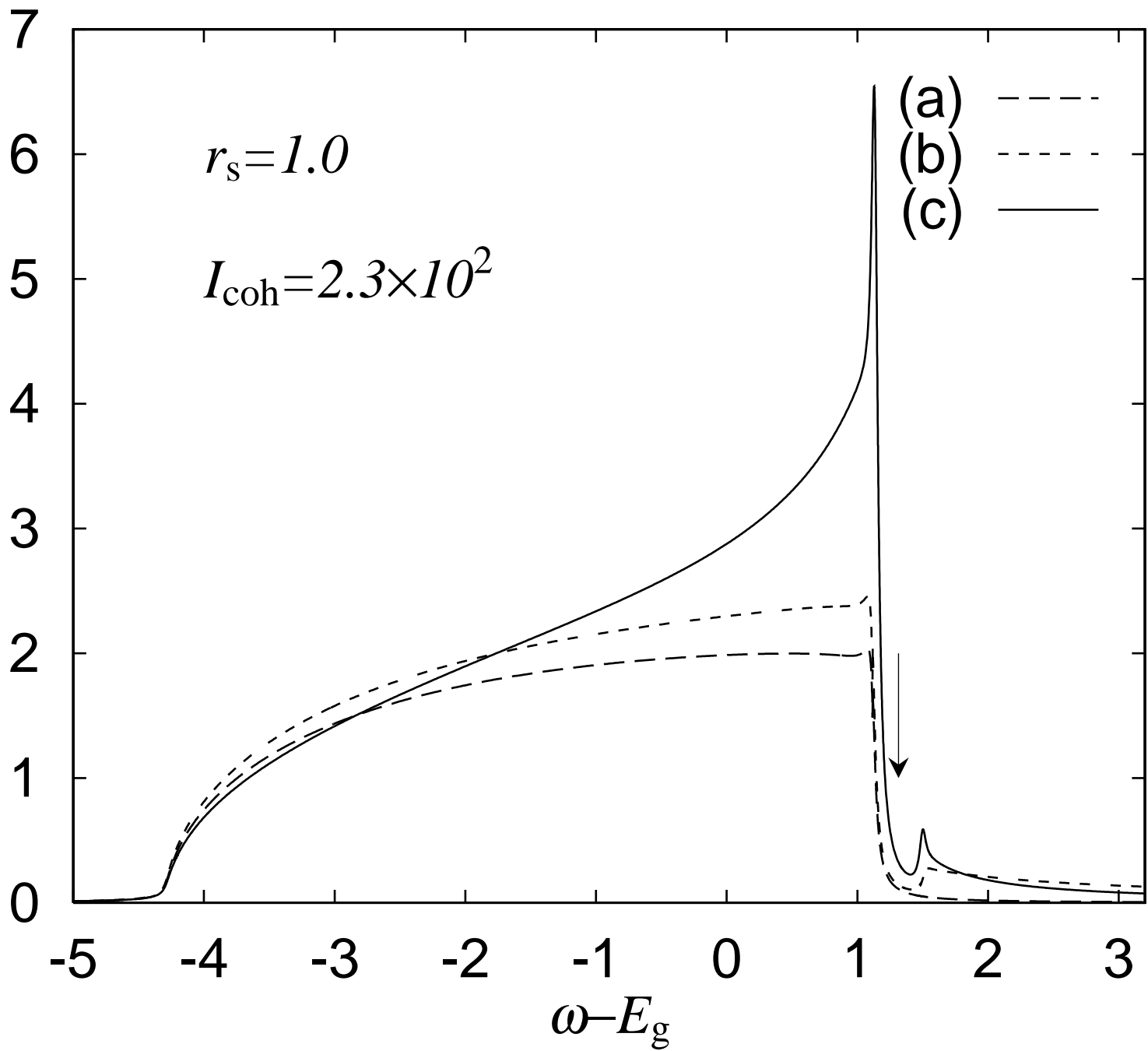
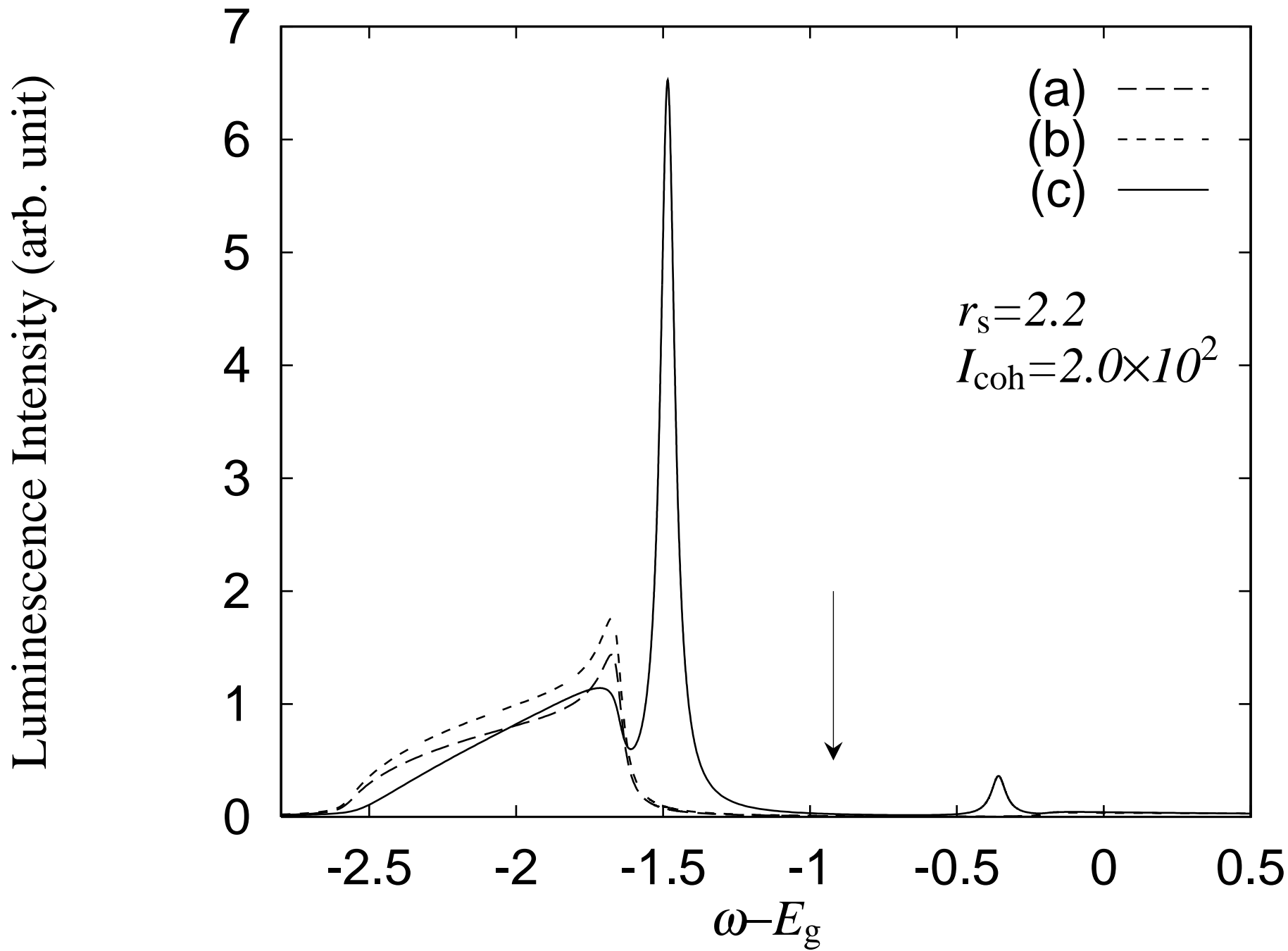
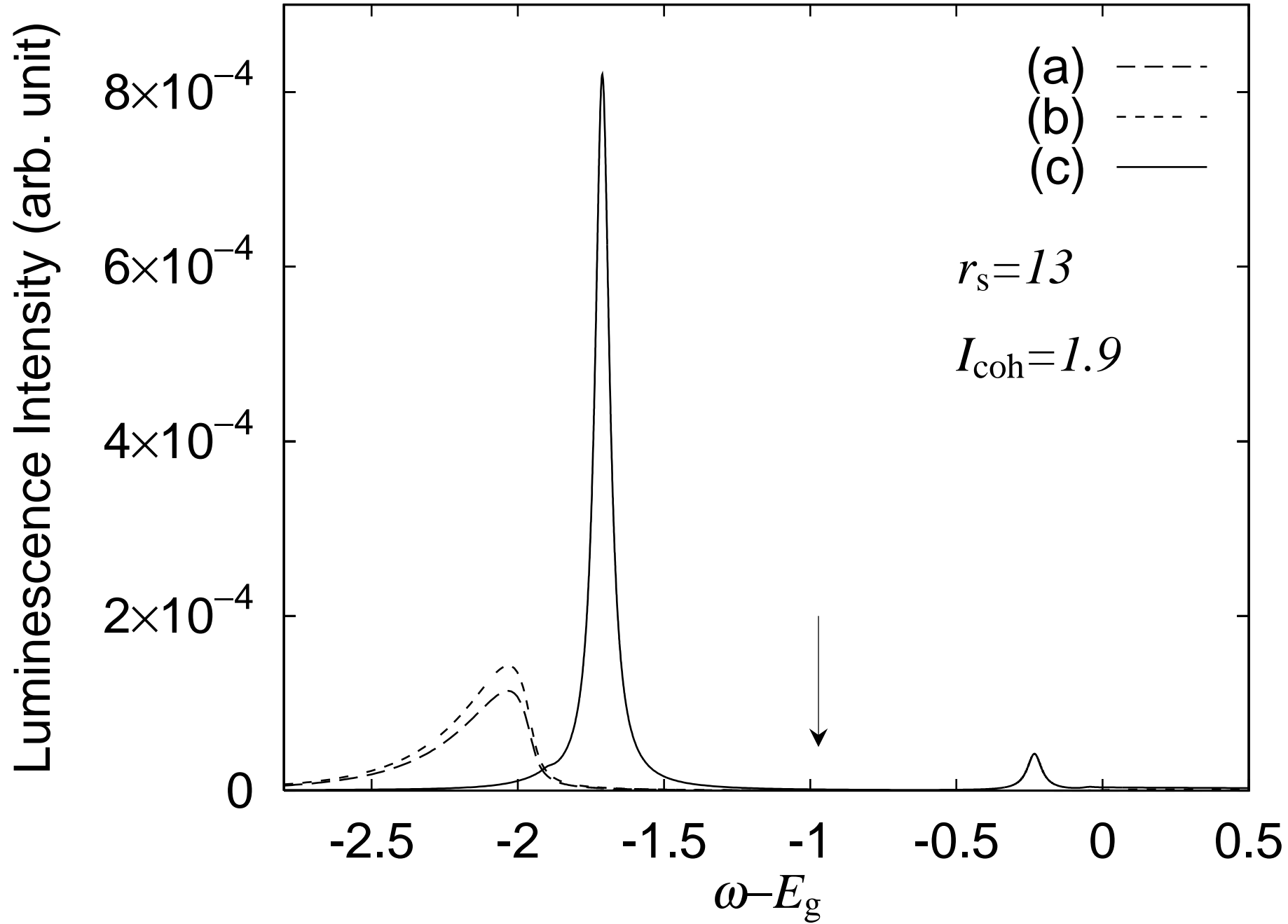
\end{document}